\documentstyle[epsbox,12pt]{article}

\begin{document}

\newcommand{\dfrac}[2]{\displaystyle{\frac{#1}{#2}}}

{\it University of Shizuoka}

\hspace*{10cm}{\bf US-98-08}\\[-.3in]

\hspace*{10cm}{\bf hep-ph/9809548}\\[-.3in]

\hspace*{10cm}{\bf Revised version}\\[-.3in]

\hspace*{10cm}{\bf February 1999}\\[.4in]

\begin{center}
{\Large\bf Universal Seesaw Mass Matrix Model}\\[.2in]
{\Large\bf and SO(10)$\times$SO(10) Unification}\\[.3in]

{\bf Yoshio Koide}\footnote{E-mail: koide@u-shizuoka-ken.ac.jp}
\\[.1in]

{ Department of Physics, University of Shizuoka} \\

{ 52-1 Yada, Shizuoka 422-8526, Japan}

\vspace{2cm}

{\large\bf Abstract}

\end{center}

\begin{quotation}
On the universal seesaw mass matrix model, which is
a promising model of the unified description of the
quark and lepton mass matrices, the behaviors of
the gauge coupling constants and intermediate 
energy scales in the ${\rm SO(10)}_L\times{\rm SO(10)}_R$ 
model are investigated related to the neutrino mass
generation scenarios.
The unification of the gauge coupling constants in the
framework of the non-SUSY model is possible if the SO(10)
symmetry is broken via Pati-Salam type symmetries.
\end{quotation}

\vfill
{Key wards: universal seesaw, evolution, SO(10), quark mass matrix, 

neutrino mass matrix}

{PACS numbers: 11.10.Hi, 12.60.-i, 14.60.Pq }

To be published in Euro.Phys.Jour.C, 1999.

\newpage

\section{Introduction}
\label{sec:1}


Recently, considerable 
interest\cite{KFzp,Morozumi,Koide-mnu,Koide-evol,recentUSMM} 
in the universal seesaw mass matrix model\cite{USMM} 
has been revived 
as a unified mass matrix model of the quarks and leptons.
Suggested by the seesaw mechanism for 
neutrinos\cite{seesaw-nu}, the model was proposed 
in order to understand the question why the masses of 
quarks (except for top quark) and charged leptons are
so small compared with the electroweak scale
$\Lambda_L$ ($\sim$ 10$^2$ GeV).
The model has hypothetical fermions
$F_i$ in addition to the conventional quarks
and leptons $f_i$ (flavors $f=u, d, \nu, e$; family
indices $i=1,2,3$), and those are assigned to 
$f_L$ = (2,1), $f_R$ = (1,2), $F_L$ = (1,1) and
$F_R$ = (1,1) of SU(2)$_L \times$ SU(2)$_R$.
The 6 $\times$ 6 mass matrix which is sandwiched
between the fields ($\overline{f}_L, \overline{F}_L$)
and ($f_R, F_R$) is given by
$$
M^{6 \times 6} =
\left( \begin{array}{cc}
0 & m_L\\
m_R & M_F
\end{array} \right) \ ,
\eqno(1.1)
$$
where $m_L$ and $m_R$ are universal for all fermion
sectors ($f=u, d, \nu, e$) and only $M_F$ have
structures dependent on the flavors $f$.
For $\Lambda_L < \Lambda_R \ll \Lambda_S$, where 
$\Lambda_L = O(m_L), \Lambda_R = O(m_R)$ and
$\Lambda_S = O(M_F)$, the $3\times 3$ mass matrix
$M_f$ for the fermions $f$ is given by the
well-known seesaw expression
$$
M_f \simeq - m_L M^{-1}_F m_R \ .
\eqno(1.2)
$$
However, after the observation\cite{t} of the heavy
top quark mass $m_t \sim \Lambda_L$, the model,
at one time, became embarrassed, because the observed
fact $m_t \sim O(m_L)$ means $O(M^{-1}_F m_R) \sim 1$.
This problem was recently solved by Fusaoka and
the author\cite{KFzp}, and later by Morozumi 
{\it et al}.\cite{Morozumi}.
If we can built a model with {\rm det}$M_F = 0$
for the up-quark sector ($F=U$), one of the fermion
masses $m(U_i)$ is zero [say, $m(U_3)=0$],
so that the seesaw mechanism does not work for
the third family, i.e., the fermions ($u_{3L}, U_{3R}$)
and ($u_{3R}, U_{3L}$) acquire masses of $O(m_L)$ and
$O(m_R)$, respectively. 
We identify $(u_{3L},U_{3_R})$ as the top quark 
$(t_L, t_R)$.
Thus,  we can understand the question
why only the top quark has a mass of the order of
$\Lambda_L$.
Of course, we can successfully describe\cite{KFzp} the
quark masses and mixings in terms of the charged
lepton masses by assuming simple structures of
$m_L, m_R$ and $M_F$.
The model also gives an interesting phenomenology
for neutrinos\cite{Koide-mnu}.

In spite of such phenomenological successes, there is a
reluctance to recognize the model, because the model 
needs extra fermions $F$.
In most unification models, there are no rooms for the 
fermions $F$.
For example, it has been found\cite{Koide-evol} that 
when the gauge symmetries 
SU(2)$_L\times$SU(2)$_R\times$U(1)$_Y\times$SU(3)$_c$ are
embedded into the Pati-Salam\cite{P-S} type unification
SO(10) $\rightarrow$ SU(2)$_L\times$SU(2)$_R\times$SU(4)$_{PS}$,
those gauge coupling constants are unified at $\mu=\Lambda_X
\simeq 6\times 10^{17}$ GeV [SU(4) is broken into 
U(1)$_Y\times$SU(3)$_c$ at $\mu=\Lambda_R\simeq 5\times 10^{12}$
GeV].
However, in the SO(10) model, there is no representation 
which offers suitable seats to the fermions 
$F_{L/R}=(1,1,4)_{L/R}$ of 
SU(2)$_L\times$SU(2)$_R\times$SU(4)$_{PS}$.
Whether we can built a unification model in which
the fermions $F$ are reasonably embedded will be 
a touchstone for the great future of the universal 
seesaw mass matrix model. 

For this problem, there is an idea \cite{so10}.
We can consider that the fermions $F_R^c$ 
($\equiv C\overline{F}_R^T$)
together with the fermions $f_L$ belong to {\bf 16} of SO(10), and 
also $F_L^c$ together with $f_R$ belong to {\bf 16} of another
SO(10), i.e.,
$$
(f_L + F_R^c) \sim (16,1)\ , \ \ \ (f_R + F_L^c) \sim (1,16) \ , 
\eqno(1.3)
$$
of SO(10)$_L\times$SO(10)$_R$.
The symmetries are broken into 
SU(2)$_L\times$SU(2)$_R\times$U(1)$_Y\times$SU(3)$_c$
at $\mu=\Lambda_S$ and the fermions $F$ have the mass term 
$\overline{F}_LM_F F_R$.

In order to examine the idea (1.3), in the present paper,
we investigate the evolution of the gauge coupling constants 
on the basis of the SO(10)$_L\times$SO(10)$_R$ model and 
estimate the intermediate energy scales $\Lambda_R$ and 
$\Lambda_S$ together with the unification energy scale $\Lambda_X$. 
Of the numerical results, especially,
we interest in the value of $\kappa \equiv \Lambda_R / \Lambda_L$,
because the value is closely related to
the neutrino mass generation scenarios as
we discuss in the next section.
The evolutions of the gauge coupling constants under
SO(10)$_L\times$SO(10)$_R$ symmetries have already been done 
by Davidson, Wali and Cho\cite{so10}, but their symmetry breaking 
patterns are somewhat 
different from that in the present model. 
We will investigate the possible intermediate energy scales
under a constraint
$\Lambda_R / \Lambda_S \simeq 0.02$ \cite{KFzp} which were derived 
 from the observed ratio $m_t/m_c$ in the new scenario of the 
 universal seesaw model\cite{KFzp,Morozumi}, where the masses $m_t$ 
 and $m_c$ are given by $m_t \sim \Lambda_L$ and 
 $m_c \sim (\Lambda_R / \Lambda_S) \Lambda_L$, respectively.

In Sec.~3, we investigate the case of the symmetry breaking
${\rm SO(10)}_L \times {\rm SO(10)}_R \rightarrow 
[{\rm SU(5)}\times {\rm U(1)}']_L \times [{\rm SU(5)}\times 
{\rm U(1)}']_R$. 
We will ruled out this case, because the results are inconsistent 
with the observed values of the gauge coupling constants 
at $\mu=m_Z$.
In Sec.~4, we investigate the case 
${\rm SO(10)}_L \times {\rm SO(10)}_R \rightarrow 
[{\rm SU(2)}\times {\rm SU(2)}'\times {\rm SU(4)}]_L \times 
[{\rm SU(2)}\times {\rm SU(2)}'\times {\rm SU(4)}]_R$.
We will conclude that the case is allowed for 
the intermediate energy scale 
$\Lambda_R\sim (10^{1}-10^{6})$ GeV
if we accept a model with $\Lambda_{XL}\neq \Lambda_{XR}$,
where $\Lambda_{XL}$ and $\Lambda_{XR}$ are the unification
scales of ${\rm SO(10)}_L$ and ${\rm SO(10)}_R$, respectively.
Finally, Sec.~5 will be devoted to the conclusions and 
remarks.


\vspace{.2in}

\section{Neutrino mass matrix}
\label{sec:2}

\vglue.1in

In the universal seesaw mass matrix model, the most general 
form of the neutrino mass matrix 
which is sandwiched between 
$(\overline{\nu}_L, \overline{\nu}_R^c, \overline{N}_L, 
\overline{N}_R^c)$ and $(\nu_L^c, \nu_R, N_L^c, N_R)^T$
is  given by
$$
M^{12\times 12}=\left(
\begin{array}{cccc}
0 & 0 & m'_L & m_L \\
0 & 0 & m_R^T & m^{\prime T}_R \\
m^{\prime T}_L & m_R & M_R & M_D \\
m_L^T & m'_R & M_D^T & M_L 
\end{array} \right) \ , 
\eqno(2.1)
$$
under the broken ${\rm SU(2)}_L\times{\rm SU(2)}_R
\times {\rm U(1)_Y}$ symmetries.
Here, we have denoted the Majorana mass terms of 
the fermions $F_L^c$ and $F_R^c$ as $M_R$ and $M_L$, 
respectively, because the fermions $F_L$ and $F_R$
are members of $(1,16^*)$ and $(16^*,1)$ of 
${\rm SO(10)}_L\times{\rm SO(10)}_R$, respectively.
The mass terms $\overline{f}_L m_L F_R$ and
$\overline{F}_L m_R f_R$ are generated, for example, 
by the Higgs scalars $(126, 1)$ and $(1, 126^*)$ of 
${\rm SO(10)}_L\times{\rm SO(10)}_R$, respectively,
while the mass terms $\overline{f}_L m'_L F_L^c$ and
$\overline{F}_R^c m'_R f_R$ must be generated by
Higgs scalars of the type $(16,16^*)$ of 
${\rm SO(10)}_L\times{\rm SO(10)}_R$.
Therefore, in the present model, we do not consider
the terms $m'_L$ and $m'_R$, i.e., we take $m'_L=m'_R=0$.
(For the special case with $m'_L\simeq m_L$ and
$m'_R\simeq m_R$, see Ref.~\cite{KF_psD}.)
Hereafter, we assume $m_L\ll m_R \ll M_F$.

Our interest is in a mass matrix for the left-handed
neutrino states $\nu_L$.
By using the seesaw approximation for the matrix (2.1),
we obtain the $6\times 6$ mass matrix for approximate
$(\nu_L^c, \nu_R)$ states
$$
M^{6\times 6}\simeq -\left(
\begin{array}{cc}
0 & m_L \\
m_R^T & 0 \\
\end{array}\right) \left(
\begin{array}{cc}
M_R & M_D \\
M_D^T & M_L \\
\end{array}\right)^{-1} \left(
\begin{array}{cc}
0 & m_R \\
m_L^T & 0 \\
\end{array}\right) 
$$
$$
=-\left(
\begin{array}{cc}
m_L M_{22}^{-1} m_L^T & m_L M_{21}^{-1} m_R \\
m_R^T M_{12}^{-1} m_L^T & m_R^T M_{11}^{-1} m_R \\
\end{array}\right) \ ,
\eqno(2.2)
$$
where
$$
\left(
\begin{array}{cc}
M_R & M_D \\
M_D^T & M_L \\
\end{array}\right)^{-1}= \left(
\begin{array}{cc}
M_{11}^{-1} & M_{12}^{-1} \\
M_{21}^{-1} & M_{22}^{-1} \\
\end{array}\right) 
\eqno(2.3)
$$
$$
M_{11}=M_R -M_D M_L^{-1} M_D^T \ ,
$$
$$
M_{22}=M_L -M_D^T M_R^{-1} M_D \ ,
\eqno(2.4)
$$
$$
M_{12}=M_{21}^T=M_D^T -M_L M_D^{-1} M_R \ .
$$
According as the cases (a) $M_L, M_R \gg M_D$, 
(b) $M_L, M_R \sim M_D$, and 
(c) $M_L, M_R \ll M_D$, we obtain the following
mass matrix for the approximate $\nu_L$ states.

 (a) The case $M_L, M_R \gg M_D$

{}From $M_{11}\simeq M_R$, $M_{22}\simeq M_L$
and $M_{12}\simeq - M_L M_D^{-1} M_R$, we obtain
$$
M^{6\times 6}\simeq \left(
\begin{array}{cc}
-m_L M_L^{-1} m_L^T & m_L M_L^{-1} M_D^T M_R^{-1} m_R \\
m_R^T M_R^{-1} M_D M_L^{-1} m_L^T & -m_R^T M_R^{-1} m_R \\
\end{array}\right) \ ,
\eqno(2.5)
$$
so that we get the mass matrix for approximate
$\nu_L$ states
$$
M(\nu_L) \simeq -m_L M_L^{-1} m_L^T \ ,
\eqno(2.6)
$$
because of $(M^{6\times 6})_{11},
(M^{6\times 6})_{22}\gg (M^{6\times 6})_{12}$.

 (b) The case $M_L, M_R \sim M_D$

We consider the case
$$
{\rm det}\left(
\begin{array}{cc}
M_R & M_D \\
M_D^T & M_L \\
\end{array}\right)\neq 0 \ .
\eqno(2.7)
$$ 
(The special case that the determinant is zero has 
been discussed in Ref. \cite{Koide-mnu}.)
Since we consider the case $m_L\ll m_R$, we can use
the seesaw approximation for the expression (2.2),
so that we obtain
$$
M(\nu_L) \simeq -m_L M_{22}^{-1} m_L^T 
+m_L M_{21}^{-1} m_R (m_R^T M_{11}^{-1} m_R)^{-1}
m_R^T M_{12}^{-1} m_L^T
$$
$$
=-m_L(M_{22}^{-1}-M_{21}^{-1}M_{11}M_{12}^{-1}) m_L^T 
=- m_L M_L^{-1} m_L^T \ ,
\eqno(2.8)
$$
where we have used the relation
$M_L=(M_{22}^{-1}-M_{21}^{-1}M_{11}M_{12}^{-1})^{-1}$
in the inverse expression of (2.3).
Thus, we obtain the expression (2.6) for the case (b), too.
Note that the $3\times 3$ mass matrix for
approximate $\nu_L$ states is almost 
independent of the structures of $M_D$ and
$M_R$ in spite of $O(M_L)\sim O(M_D)\sim O(M_R)$.

 (c) The case $M_L, M_R \ll M_D$

{}From $M_{11}\simeq -M_D M_L^{-1} M_D^T$,
$M_{22}\simeq -M_D^T M_R^{-1} M_D$ and
$M_{12}\simeq M_D^T$, we obtain the mass matrix 
$$
M^{6\times 6}\simeq \left(
\begin{array}{cc}
m_L M_D^{-1} M_R M_D^{T-1} m_L^T & -m_L M_D^{-1} m_R \\
-m_R M_D^{T-1} m_L^T & m_R^T M_D^{T-1}M_L M_D^{-1} m_R \\
\end{array}\right) \ .
\eqno(2.9)
$$
The mass matrix gives three light pseudo-Dirac neutrino
states\cite{ps-D} $\nu_{i\pm}^{psD}\simeq (\nu_{iL}\pm 
\nu_{iR}^c)/\sqrt{2}$ ($i=e,\mu,\tau$), because
$(M^{6\times 6})_{11},(M^{6\times 6})_{22}\ll 
(M^{6\times 6})_{12}$.
This case has been discussed by Bowes and 
Volkas\cite{Bowes}.
The case is very attractive phenomenologically,
because the maximal mixing state between $\nu_{\mu L}$ 
and $\nu_{\mu R}$ can give a natural explanation for
the recent atmospheric neutrino data\cite{nu-atm}.
The mass matrix  $M(\nu_{\pm}^{psD})$ in the limit of 
$m(\nu_{i+}^{psD})=m(\nu_{i-}^{psD})$ is approximately
given by
$$
M(\nu_{\pm}^{psD}) \simeq - m_L M_D^{-1} m_R \ .
\eqno(2.10)
$$

First, we suppose the following symmetry breaking pattern
(hereafter, we will refer to it as the case A):
$$
{\rm SO(10)}_L \times {\rm SO(10)}_R
$$
$$
\downarrow \ \ \ \mu=\Lambda_{X10}
$$
$$
[{\rm SU(5)}\times {\rm U(1)}']_L \times [{\rm SU(5)}\times 
{\rm U(1)}']_R
$$
$$
\downarrow \ \ \ \mu=\Lambda_{N}
$$
$$
{\rm SU(5)}_L \times {\rm SU(5)}_R
$$
$$
\downarrow \ \ \ \mu=\Lambda_{X5}
$$
$$
[{\rm SU(3)}\times {\rm SU(2)}\times {\rm U(1)}]_L \times
[{\rm SU(3)}\times {\rm SU(2)}\times {\rm U(1)}]_R  
$$
$$
\downarrow \ \ \ \mu=\Lambda_{S}
$$
$$
{\rm SU(3)}_c \times {\rm SU(2)}_L\times {\rm SU(2)}_R
\times {\rm U(1)}_Y 
$$
$$
\downarrow \ \ \ \mu=\Lambda_{R}
$$
$$
{\rm SU(3)}_c \times {\rm SU(2)}_L\times {\rm U(1)}_{Y'}
$$
$$
\downarrow \ \ \ \mu=\Lambda_{L}
$$
$$
{\rm SU(3)}_c \times {\rm U(1)}_{em} \ .
\eqno(2.11)
$$
At the energy scale $\mu=\Lambda_N$, 
the gauge symmetries ${\rm U(1)'}_L \times {\rm U(1)'}_R$ 
are completely broken, so that the neutral leptons 
$N_L$ and $N_R$ acquire Dirac and Majorana masses
of the order of $\Lambda_N$.
At $\mu=\Lambda_S$, the remaining fermions $F_L$ and
$F_R$ (except for $U_{3L}$ and $U_{3R}$) acquire masses 
of the order of $\Lambda_S$ by Higgs bosons
$\Phi$ (as we discuss in the next section),
and ${\rm SU(3)}_L \times {\rm SU(3)}_R$ and 
${\rm U(1)}_L \times {\rm U(1)}_R$ are broken into
${\rm SU(3)}_{L+R}\equiv {\rm SU(3)}_c$ and
${\rm U(1)}_{L+R}\equiv {\rm U(1)}_Y$, respectively.
If this scenario A is true, the mass matrices
$M_L, M_R$ and $M_D$ are of the order of
$\Lambda_N$, so that we suppose that the order of
the neutrino masses $m(\nu_i)$ are given by
$$
m(\nu_i) \sim \Lambda_L^2 / \Lambda_L \sim
(\Lambda_L \Lambda_S / \Lambda_R \Lambda_N)
m(e_i)\ ,
\eqno(2.12)
$$
from the result (2.8) in the case (b),
the neutrino masses are suppressed by a factor
($\Lambda_L / \Lambda_R$) ($\Lambda_S / \Lambda_N$)
compared with the charged lepton masses $m(e_i)$.

Next, we can suppose another symmetry breaking (case B):
$$
{\rm SO(10)}_L \times {\rm SO(10)}_R
$$
$$
\downarrow \ \ \ \mu=\Lambda_{X}
$$
$$
[{\rm SU(2)}\times{\rm SU(2)'}\times{\rm SU(4)}]_L \times 
[{\rm SU(2)}\times{\rm SU(2)'}\times{\rm SU(4)}]_R
$$
$$
\downarrow \ \ \ \mu=\Lambda_{S}
$$
$$
{\rm SU(2)}_L\times {\rm SU(2)}_R\times {\rm U(1)}_Y 
\times {\rm SU(3)}_c  
$$
$$
\downarrow \ \ \ \mu=\Lambda_{R}
$$
$$
{\rm SU(3)}_c \times {\rm SU(2)}_L\times {\rm U(1)}_{Y'}
$$
$$
\downarrow \ \ \ \mu=\Lambda_{L}
$$
$$
{\rm SU(3)}_c \times {\rm U(1)}_{em} \ .
\eqno(2.13)
$$
If this scenario B is true, since
$M_L \sim M_R \sim M_D \sim M_S$, we suppose
$$
m(\nu_i) \sim \Lambda_L^2 / \Lambda_S \sim
(\Lambda_L / \Lambda_R) m(e_i),
\eqno(2.14)
$$
so that the neutrino masses $m(\nu_i)$ are suppressed by
a factor $\Lambda_L / \Lambda_R$ compared with
the charged lepton masses $m(e_i)$.

What is of the great interest is
to estimate the possible values of such intermediate
energy scales $\Lambda_R$, $\Lambda_S$, and so on.

Although the Bowes-Volkas model\cite{Bowes} is very 
interesting,  the model cannot
apply to the universal seesaw model based on the 
${\rm SO(10)}_L\times {\rm SO(10)}_R$ unification, 
because the case $M_L, M_R\ll M_D$ is not likely in the
${\rm SO(10)}_L\times {\rm SO(10)}_R$ model, and, if it is
possible, the relation (2.10) leads to the wrong prediction
$m(\nu_i) \sim m(e_i)$ for $M_D \equiv M_N \sim M_F$   
($F \neq N$). 


\vspace{.2in}

\section{Case of ${\rm SO(10)}\rightarrow {\rm SU(5)}\times{\rm U(1)}$}
\label{sec:3}


In the present section, we investigate
the case A with the symmetry breaking pattern (2.11).
At the energy scale $\mu=\Lambda_S$,
the symmetries $[{\rm SU(3)} \times {\rm SU(2)} \times {\rm U(1)}]_L 
\times [{\rm SU(3)} \times SU(2) \times U(1)]_R$ are
broken into ${\rm SU(3)}_c \times {\rm SU(2)}_L \times
{\rm U(1)}_Y$ by the following Higgs scalars $\Phi_Y$:
$$
\Phi_{2/3} \sim (3^*,1;3,1)_{Y=2/3},
$$
$$
\Phi_{4/3} \sim (3,1;3^*,1)_{Y=4/3},
\eqno(3.1)
$$
$$
\Phi_2 \sim (1,1;1,1)_{Y=2},
$$
of [${\rm SU(3)} \times {\rm SU(2)}]_L \times
[{\rm SU(3)} \times {\rm SU(2)}]_R$, where
${\rm SU(3)}_c \equiv {\rm SU(3)}_{L+R},
{\rm U(1)}_Y \equiv {\rm U(1)}_{L+R}$,
and $Y = Y_L = Y_R$.
Our interest is in the region
$\Lambda_L < \mu \leq \Lambda_{X5}$.
Hereafter, we call region $\Lambda_L < \mu \leq \Lambda_R$, 
$\Lambda_R < \mu \leq \Lambda_S$,
and $\Lambda_S < \mu \leq \Lambda_{X5}$
regions I, II, and III, respectively.

The electric charge operator $Q$ is given by
$$
Q=I_3^L +\frac{1}{2} Y' 
\ \ \ \ {\rm (Region\ I)}\ ,
\eqno(3.2)
$$
$$
\frac{1}{2} Y'=I_3^R + \frac{1}{2} Y 
\ \ \ \ {\rm (Region\ II)}\ ,
\eqno(3.3)
$$
$$
\frac{1}{2} Y= \frac{1}{2} Y_L +\frac{1}{2} Y_R 
\ \ \ \ {\rm (Region\ III)}\ .
\eqno(3.4)
$$
We denote the gauge coupling constants corresponding to the
operators $Q$, $Y'$, $Y$, $Y_L$, $Y_R$, $I^L$, and $I^R$ as
$g_{em}\equiv e$, $g'_1$, $g_1$, $g_{1L}$, $g_{1R}$, $g_{2L}$,
and $g_{2R}$, respectively.
The boundary conditions for these gauge coupling constants
at $\mu=\Lambda_L$, $\mu=\Lambda_R$, and $\mu=\Lambda_S$ 
are as follows:
$$
\alpha^{-1}_{em}(\Lambda_L) = \alpha_{2L}^{-1}(\Lambda_L)+
\frac{5}{3}\alpha_1^{\prime -1}(\Lambda_L) \ ,
\eqno(3.5)
$$
$$
\frac{5}{3}\alpha^{\prime -1}_{1}(\Lambda_R) = 
\alpha_{2R}^{-1}(\Lambda_R)+
\frac{2}{3}\alpha_1^{-1}(\Lambda_R) \ ,
\eqno(3.6)
$$
and 
$$
\frac{2}{3}\alpha^{-1}_{1}(\Lambda_S) = 
\frac{5}{3}\alpha^{-1}_{1L}(\Lambda_S)+
\frac{5}{3}\alpha_{1R}^{-1}(\Lambda_S) \ ,
\eqno(3.7)
$$
respectively, correspondingly to Eqs.~(3.2), (3.3) and 
(3.4), where  $\alpha_i\equiv g_i^2/4\pi$ and 
the normalizations of the U(1)$_{Y'}$, U(1)$_Y$, 
U(1)$_{Y_L}$ and U(1)$_{Y_R}$ gauge 
coupling constants have been taken as they satisfy 
$\alpha'_1=\alpha_{2L}=\alpha_3$, 
$\alpha_{1L}=\alpha_{2L}=\alpha_{3L}$, and  
$\alpha_{1R}=\alpha_{2R}=\alpha_{3R}$  in the SU(5) 
grand-unification limit and 
$\alpha_1=\alpha_3\equiv \alpha_4$ in the SU(4) 
unification limit [$\alpha_4=\alpha_{2L}=\alpha_{2R}$ 
in the SO(10) unification limit], respectively.
We also have the following boundary conditions at 
$\mu=\Lambda_S$ and $\mu=\Lambda_{X5}$:
$$
\alpha^{-1}_{3}(\Lambda_S) = \alpha^{-1}_{3L}(\Lambda_S)+
\alpha_{3R}^{-1}(\Lambda_S) \ ,
\eqno(3.8)
$$
$$
\alpha^{-1}_{1L}(\Lambda_{X5L}) = \alpha^{-1}_{2L}
(\Lambda_{X5L}) = \alpha_{3L}^{-1}(\Lambda_{X5L}) \ ,
\eqno(3.9)
$$
$$
\alpha^{-1}_{1R}(\Lambda_{X5R}) = \alpha^{-1}_{2R}
(\Lambda_{X5R}) = \alpha_{3R}^{-1}(\Lambda_{X5R}) \ ,
\eqno(3.10)
$$
where, for convenience, we distinguish the unification
scale of ${\rm SU(5)}_L$, $\Lambda_{5XL}$, from that
of ${\rm SU(5)}_R$, $\Lambda_{X5R}$.

The evolutions of the gauge coupling constants $g_i$ at
one-loop are given by the equations
$$
\frac{d}{dt} \alpha_i(\mu) = -\frac{1}{2\pi}
 b_i \alpha_i^2(\mu) \  , 
\eqno(3.11)
$$
where $t=\ln \mu$.
Since the quantum numbers of the fermions $f$ and $F$ are 
assigned as those in Table \ref{T-qn},  the coefficients 
$b_i$ are given in Table \ref{T-b}.
In the model with det$M_U=0$, the heavy fermions $F_L$ and 
$F_R$ except for $U_{3L}$ and $U_{3R}$ are decoupled for 
$\mu\leq\Lambda_S$ and the fermions $u_{3R}$ and $U_{3L}$ 
are decoupled for $\mu\leq\Lambda_R$.
In Table \ref{T-b}, we have also shown the values of $b_i$
for the conventional case without such the constraint
det$M_U=0$ in parentheses.

By substituting $\alpha^{-1}_{2L}(\Lambda_{X5L})=
\alpha^{-1}_{3L}(\Lambda_{X5L})$ with the relations
at one-loop
$$
\alpha^{-1}_{2L}(\Lambda_{X5L}) = \alpha^{-1}_{2L}
(\Lambda_{S}) + b^{III}_{2L}\frac{1}{2\pi}\ln 
\frac{\Lambda_{X5L}}{\Lambda_S} \ , 
\eqno(3.12)
$$
$$
\alpha^{-1}_{3L}(\Lambda_{X5L}) = \alpha^{-1}_{3L}
(\Lambda_{S}) + b^{III}_{3L}\frac{1}{2\pi}\ln 
\frac{\Lambda_{X5L}}{\Lambda_S} \ , 
\eqno(3.13)
$$
we obtain
$$
\alpha^{-1}_{3L}(\Lambda_{S})-\alpha^{-1}_{2L}(\Lambda_{S})
 + (b^{III}_{3L}-b^{III}_{2L})\frac{1}{2\pi}\ln 
\frac{\Lambda_{X5L}}{\Lambda_S}=0 \ . 
\eqno(3.14)
$$
Similarly, from the condition 
$\alpha^{-1}_{1L}(\Lambda_{X5L})=
\alpha^{-1}_{2L}(\Lambda_{X5L})$, we obtain
$$
\alpha^{-1}_{2L}(\Lambda_{S})-\alpha^{-1}_{1L}(\Lambda_{S})
 + (b^{III}_{2L}-b^{III}_{1L})\frac{1}{2\pi}\ln 
\frac{\Lambda_{X5L}}{\Lambda_S}=0 \ . 
\eqno(3.15)
$$
By eliminating $\ln(\Lambda_{X5L}/\Lambda_S)$ from Eqs.~(3.14)
and (3.15), we obtain
$$
(b^{III}_{2L}-b^{III}_{1L})\alpha^{-1}_{3L}(\Lambda_{S})
+ (b^{III}_{3L}-b^{III}_{2L})\alpha^{-1}_{1L}(\Lambda_{S})
-(b^{III}_{3L}-b^{III}_{1L})\alpha^{-1}_{2L}(\Lambda_{S})=0
\ .
\eqno(3.16)
$$
Similarly, we obtain
$$
(b^{III}_{2R}-b^{III}_{1R})\alpha^{-1}_{3R}(\Lambda_{S})
+ (b^{III}_{3R}-b^{III}_{2R})\alpha^{-1}_{1R}(\Lambda_{S})
-(b^{III}_{3R}-b^{III}_{1R})\alpha^{-1}_{2R}(\Lambda_{S})=0
\ .
\eqno(3.17)
$$
Therefore, from the relations (3.7), (3.8) and 
$b^{III}_{iL}=b^{III}_{iR}\equiv b^{III}_{i}$, we obtain
$$
(b^{III}_{2}-b^{III}_{1})\alpha^{-1}_{3}(\Lambda_{S})
+ (b^{III}_{3}-b^{III}_{2})\alpha^{-1}_{1}(\Lambda_{S})
$$
$$
-(b^{III}_{3}-b^{III}_{1})\left[\alpha^{-1}_{2L}(\Lambda_{S})
+\alpha^{-1}_{2R}(\Lambda_{S})\right]=0
\ ,
\eqno(3.18)
$$
which leads to
$$
\left[\frac{3}{5}(b^{III}_3-b^{III}_2)+
(b^{III}_3-b^{III}_1)\right]\alpha^{-1}_{2R}(\Lambda_{R})
$$
$$
-\left[b^{II}_3(b^{III}_2-b^{III}_1)+\frac{2}{5}
b^{II}_1(b^{III}_3-b^{III}_2)-2
b^{II}_2(b^{III}_3-b^{III}_1)\right]
\frac{1}{2\pi}\ln \frac{\Lambda_{S}}{\Lambda_R}
$$
$$
-\left[b^{I}_3(b^{III}_2-b^{III}_1)+
b^{I}_1(b^{III}_3-b^{III}_2)-
b^{I}_2(b^{III}_3-b^{III}_1)\right]
\frac{1}{2\pi}\ln \frac{\Lambda_{R}}{\Lambda_L}
$$
$$
=(b^{III}_{2}-b^{III}_{1})\alpha^{-1}_{3}(\Lambda_{L})
+ (b^{III}_{3}-b^{III}_{2})\alpha^{\prime -1}_{1}
(\Lambda_{L})
-(b^{III}_{3}-b^{III}_{1})\alpha^{-1}_{2L}(\Lambda_{L})
\ .
\eqno(3.19)
$$
For the model with det$M_U=0$, the relation (3.19) becomes
$$
13 \alpha^{-1}_{2R}(\Lambda_{R})+
\frac{391}{15}\frac{1}{2\pi}\ln \frac{\Lambda_{S}}{\Lambda_R}
- \frac{178}{15} \frac{1}{2\pi}\ln \frac{\Lambda_R}{\Lambda_L}
$$
$$
=\frac{127}{15}\alpha^{-1}_{3}(\Lambda_{L})
+ \frac{17}{6}\alpha^{\prime -1}_{1}(\Lambda_{L})
-\frac{113}{30}\alpha^{-1}_{2L}(\Lambda_{L})
\ .
\eqno(3.20)
$$
The right-hand side of (3.20) gives the value $-97.82$ for
the input values $\alpha'_1(m_Z)=0.01683$, 
$\alpha_L(m_Z)=0.03349$ and $\alpha_3(m_Z)=0.1189$ \cite{pdg}, 
where for convenience, we have used the initial values at 
$\mu=m_Z$ instead of those at $\mu=\Lambda_L$.
The relation (3.20) puts a lower bound on the ratio
$\Lambda_R / \Lambda_L$:
For $\alpha_{2R}^{-1} (\Lambda_R) \geq$ 1,
we obtain $\Lambda_R / \Lambda_L \geq 2\times 10^{135}$
(for $\Lambda_S / \Lambda_R = 50$ \cite{KFzp})
and $\Lambda_R / \Lambda_L \geq 3\times 10^{22}$
(for $\Lambda_S / \Lambda_R \geq 1$).
Such a large value of $\Lambda_R / \Lambda_L$ is
physically unlikely, so that the case A is ruled out.

By similar discussion to the relation (3.19), it turn
out that the conclusion that the case A is ruled out 
is still unchanged for the model without the condition 
det$M_U=0$ and also for the minimal SUSY version of 
the present model.


\vspace{.2in}

\section{Case of ${\rm SO(10)}\rightarrow {\rm SU(2)}\times
{\rm SU(2)}\times{\rm SU(4)}$}
\label{sec:4}


Next, we investigate the case B, ${\rm SO(10)}_L \times
{\rm SO(10)}_R \rightarrow [{\rm SU(2)} \times {\rm SU(2)}' 
\times {\rm SU(4)}]_L \times [{\rm SU(2)}\times{\rm SU(2)}'
\times {\rm SU(4)}]_R$.
At the energy scale $\mu=\Lambda_S$, the symmetries
$[{\rm SU(2)}'\times{\rm SU(4)}]_L \times 
[{\rm SU(2)}'\times{\rm SU(4)}]_R$ are broken into
${\rm U(1)}_Y \times {\rm SU(3)}_c$ by  Higgs scalars
$$
\begin{array}{l}
\Phi_V \sim (1,2,4;1,2,4) , \\
\Phi_L\sim (1,1,10;1,1,1) , \\
\Phi_R \sim (1,1,1;1,1,10) ,
\end{array}
\eqno(4.1)
$$
of $[{\rm SU(2)} \times {\rm SU(2)}' \times {\rm SU(4)}]_L 
\times [{\rm SU(2)}\times{\rm SU(2)}'\times {\rm SU(4)}]_R$,
where Higgs scalars $\Phi_V$, $\Phi_L$ and $\Phi_R$ generate 
the masses $M_F$, $M_L$ and $M_R$, respectively.
In the present section, we call the regions 
$\Lambda_L <\mu\leq \Lambda_R$, 
$\Lambda_R <\mu\leq \Lambda_S$, and 
$\Lambda_S <\mu\leq \Lambda_{X}$ 
regions I, II, and III, respectively.

The electric charge operator $Q$ is given by
Eqs.~(3.2) and (3.3) in the regions I and II, 
respectively, but the relation (3.4) is replaced by
$$
\frac{1}{2} Y=I_{3}^{\prime L}+ \frac{1}{2} Y_L 
+I_{3}^{\prime R} +\frac{1}{2} Y_R \ ,
\eqno(4.2)
$$
so that the boundary condition (3.7)
is replaced by
$$
\frac{2}{3}\alpha^{-1}_{1}(\Lambda_S) = 
\alpha^{\prime -1}_{2L}(\Lambda_S)+\frac{2}{3}\alpha^{-1}_{1L}(\Lambda_S)
+\alpha^{\prime -1}_{2R}(\Lambda_S)+\frac{2}{3}\alpha_{1R}^{-1}(\Lambda_S) \ .
\eqno(4.3)
$$
The boundary conditions at $\mu=\Lambda_S$ and $\mu=\Lambda_{X}$
are as follows:
$$
\alpha^{-1}_{3}(\Lambda_S) = \alpha^{-1}_{3L}(\Lambda_S)+
\alpha_{3R}^{-1}(\Lambda_S) \ ,
\eqno(4.4)
$$
$$
\alpha^{-1}_{1L}(\Lambda_{S}) = \alpha^{-1}_{3L}
(\Lambda_{S}) = \alpha_{4L}^{-1}(\Lambda_{S}) \ ,
\eqno(4.5)
$$
$$
\alpha^{-1}_{1R}(\Lambda_{S}) = \alpha^{-1}_{3R}
(\Lambda_{S}) = \alpha_{4R}^{-1}(\Lambda_{S}) \ ,
\eqno(4.6)
$$
$$
\alpha^{-1}_{2L}(\Lambda_{XL}) = \alpha^{\prime -1}_{2L}
(\Lambda_{XL}) = \alpha_{4L}^{-1}(\Lambda_{XL}) \ ,
\eqno(4.7)
$$
$$
\alpha^{-1}_{2R}(\Lambda_{XR}) = \alpha^{\prime -1}_{2R}
(\Lambda_{XR}) = \alpha_{4R}^{-1}(\Lambda_{XR}) \ ,
\eqno(4.8)
$$
where, for convenience, we have again distinguished the
unification scale of SO(10)$_L$, $\Lambda_{XL}$,
from that of SO(10)$_R$, $\Lambda_{XR}$.

Since $b^{\prime III}_{2L}=b^{\prime III}_{2R}
\equiv b^{\prime III}_{2} \neq b^{III}_{2L}
=b^{III}_{2R} \equiv b^{III}_2$,
we obtain
$$
\alpha^{\prime -1}_{2L}(\Lambda_S) - \alpha^{-1}_{2L}(\Lambda_S)
= (b^{\prime III}_{2L} - b^{III}_{2L}) \frac{1}{2\pi} \ln
\frac{\Lambda_S}{\Lambda_{XL}},
\eqno(4.9)
$$
$$
\alpha^{\prime -1}_{2R}(\Lambda_S) = \alpha^{-1}_{2R}(\Lambda_S)
= (b^{\prime III}_{2R} - b^{III}_{2R}) \frac{1}{2\pi} \ln
\frac{\Lambda_S}{\Lambda_{XR}},
\eqno(4.10)
$$
i.e.,
$$
\alpha^{\prime -1}_{2L} (\Lambda_S) + \alpha^{\prime -1}_{2R} (\Lambda_S)
$$
$$
= \alpha^{-1}_{2L} (\Lambda_S) + \alpha^{-1}_{2R} (\Lambda_S)
+ 2(b^{III}_2 - b^{\prime III}_2)
\frac{1}{2\pi} \ln \frac{\Lambda_X}{\Lambda_S},
\eqno(4.11)
$$
where $\Lambda_X = (\Lambda_{XL} \Lambda_{XR})^{1/2}$.
On the other hand, from Eqs.(4.3)-(4.6),
we obtain
$$
\alpha^{-1}_{3}(\Lambda_{S})
+\frac{3}{2}\left[\alpha^{\prime -1}_{2L}(\Lambda_{S})
+\alpha^{\prime -1}_{2R}(\Lambda_{S})\right]
-\alpha^{-1}_{1}(\Lambda_{S}) =0
 \ , 
\eqno(4.12)
$$
so that
$$
\alpha^{-1}_3 (\Lambda_S)
+ \frac{3}{2} \left[\alpha^{-1}_{2L} (\Lambda_S)
+ \alpha^{-1}_{2R} (\Lambda_S)\right] - \alpha^{-1}_1 (\Lambda_S)
+ 3(b^{III}_2 - b^{\prime III}_2)
\frac{1}{2\pi} \ln \frac{\Lambda_X}{\Lambda_S} = 0.
\eqno(4.13)
$$

Similarly, from Eq.(4.7), we obtain
$$
\alpha^{-1}_{3L} (\Lambda_S) - \alpha^{-1}_{2L}(\Lambda_S)
+ (b^{III}_{4L} - b^{III}_{2L})
\frac{1}{2\pi} \ln \frac{\Lambda_{XL}}{\Lambda_S} = 0,
\eqno(4.14)
$$
so that, together with the equation with (L $\rightarrow$ R)
in (4.14), we obtain
$$
\alpha^{-1}_3 (\Lambda_S) - \left[\alpha^{-1}_{2L} (\Lambda_S)
+ \alpha^{-1}_{2R} (\Lambda_S)\right] + 2(b^{III}_4 - b^{III}_2)
\frac{1}{2\pi} \ln \frac{\Lambda_X}{\Lambda_S} = 0.
\eqno(4.15)
$$
By eliminating $\Lambda_X / \Lambda_R$ from (4.13) and
(4.15), we obtain
$$
c_3 \ \alpha^{-1}_3 (\Lambda_S) + c_2 \left[\alpha^{-1}_{2L}
(\Lambda_S) + \alpha^{-1}_{2R} (\Lambda_S)\right]
- c_1 \ \alpha^{-1}_1 (\Lambda_S) = 0,
\eqno(4.16)
$$
where
$$
c_1 = b^{III}_4 - b^{III}_2,
\eqno(4.17)
$$
$$
c_2 = \frac{3}{2} (b^{III}_4 - b^{\prime III}_2),
\eqno(4.18)
$$
$$
c_3 = b^{III}_4 - b^{III}_2 - \frac{3}{2} (b^{III}_2 - b^{\prime III}_2).
\eqno(4.19)
$$
Since
$$
\alpha^{-1}_1 (\Lambda_S) = \frac{5}{2} \alpha^{-1}_1 (\Lambda_L)
- \frac{3}{2} \alpha^{-1}_{2R} (\Lambda_R)
+ b^{II}_1 \frac{1}{2\pi} \ln \frac{\Lambda_S}{\Lambda_R}
+ \frac{5}{2} b^{II}_1 \frac{1}{2\pi} \ln \frac{\Lambda_R}{\Lambda_L},
\eqno(4.20)
$$
$$
\alpha^{-1}_{2L} (\Lambda_S) = \alpha^{-1}_{2L} (\Lambda_L)
+ b^{II}_2 \frac{1}{2\pi} \ln \frac{\Lambda_S}{\Lambda_R}
+ b^I_2 \frac{1}{2\pi} \ln \frac{\Lambda_R}{\Lambda_L},
\eqno(4.21)
$$
$$
\alpha^{-1}_{2R} (\Lambda_S) = \alpha^{-1}_{2L} (\Lambda_R)
+ b^{II}_2 \frac{1}{2\pi} \ln \frac{\Lambda_S}{\Lambda_R},
\eqno(4.22)
$$
$$
\alpha^{-1}_3 (\Lambda_S) = \alpha^{-1}_3 (\Lambda_L)
+ b^{II}_3 \frac{1}{2\pi} \ln \frac{\Lambda_S}{\Lambda_R}
+ b^I_3 \frac{1}{2\pi} \ln \frac{\Lambda_R}{\Lambda_L},
\eqno(4.23)
$$
the relation (4.16) leads to the constraint for
$\Lambda_R / \Lambda_L$:
$$
0 = \left(c_2 + \frac{3}{2} c_1\right) \alpha^{-1}_{2R} (\Lambda_R)
+ (c_3 b^{II}_3 + 2 c_2 b^{II}_2 - c_1 b^{II}_1)
\frac{1}{2\pi} \ln \frac{\Lambda_S}{\Lambda_R}
$$
$$
+\left(c_3 b^I_3 + c_2 b^I_2 - \frac{5}{2} b^I_1\right)
\frac{1}{2\pi} \ln \frac{\Lambda_R}{\Lambda_L}
+ c_3 \alpha^{-1}_3 (\Lambda_L) + c_2 \alpha^{-1}_{2L}
(\Lambda_L) - \frac{5}{2} \alpha^{-1}_1 (\Lambda_L)
$$
$$
= 19.5\, \alpha^{-1}_{2R} (\Lambda_R)
+ 19.67 \, \log \frac{\Lambda_R}{\Lambda_L}
+ 32.31 \, \log \frac{\Lambda_S}{\Lambda_R} - 193.96 \ ,
\eqno(4.24)
$$
where we have used the values of $b_i$ given
in Table 2 and the same input values of
$\alpha^{-1}_1 (\Lambda_L)$, $\alpha^{-1}_2 (\Lambda_L)$
and $\alpha^{-1}_3 (\Lambda_L)$ as those used in (3.20).
For $\Lambda_S / \Lambda_R = 50$, the relation (4.24)
leads to
$$
\log \frac{\Lambda_R}{\Lambda_L} = 7.071 - 0.9915\, 
\alpha^{-1}_{2R} (\Lambda_R) \ ,
\eqno(4.25)
$$
so that, for $\alpha^{-1}_{2R} (\Lambda_R) \geq 1$,
we obtain the constraint
$$
\kappa\equiv \Lambda_R / \Lambda_L \leq 1.20 \times 10^6\ .
\eqno(4.26)
$$
Similarly, we can obtain the constraint for
$\Lambda_X / \Lambda_S$:
$$
\log \frac{\Lambda_X}{\Lambda_S} = 4.098 + 0.8517\, 
\alpha^{-1}_{2R} (\Lambda_R) \ .
\eqno(4.27)
$$
We show the values of $\Lambda_R$, $\Lambda_S$ and $\Lambda_X$
for the typical values of $\alpha^{-1}_{2R} (\Lambda_R)$
in Table 3.
The values of $\Lambda_{XL}$ and $\Lambda_{XR}$ depend not only
on the input value of $\alpha^{-1}_{2R}(\Lambda_R)$ but also
on that of $\alpha^{-1}_{4R}(\Lambda_S)$, because
$$
\alpha^{-1}_{4R}(\Lambda_S)=\alpha^{-1}_{2R}(\Lambda_S)
+(b_2^{III}-b_4^{III})\frac{1}{2\pi}
\ln\frac{\Lambda_{XR}}{\Lambda_{S}}
$$
$$
=\frac{1}{2}\left[\alpha^{-1}_{2R}(\Lambda_S)-
\alpha^{-1}_{2L}(\Lambda_S)+\alpha^{-1}_{3}(\Lambda_S)
\right] +(b_4^{III}-b_4^{III})\frac{1}{2\pi}
\ln\frac{\Lambda_{X}}{\Lambda_{XR}}
$$
$$
=-3.785 -0.1964\, \alpha^{-1}_{2R}(\Lambda_R)
+1.405\, \log\frac{\Lambda_{X}}{\Lambda_{XR}} \ ,
\eqno(4.28)
$$
i.e.,
$$
\log\frac{\Lambda_{X}}{\Lambda_{XR}}=2.694
+0.1398\, \alpha^{-1}_{2R}(\Lambda_R)
+0.7118\, \alpha^{-1}_{4R}(\Lambda_S) \ ,
\eqno(4.29)
$$
where we have used $\Lambda_S/\Lambda_R=50$.
For $\alpha^{-1}_{2R}(\Lambda_R)\geq 1$ and
$\alpha^{-1}_{4R}(\Lambda_S)\geq 1$, the 
relation (4.29) gives the constraint
$$
\Lambda_{XL}/\Lambda_{XR}\geq 1.26\times 10^7 \ .
\eqno(4.30)
$$
The relation (4.29) concludes that a model with
$\Lambda_{XL} = \Lambda_{XR}$ is ruled out.
Values of $\Lambda_{XR}$ and $\Lambda_{XL}$
for typical values of $\alpha^{-1}_{2R}(\Lambda_R)$ and
$\alpha^{-1}_{4R} (\Lambda_S)$ are
also listed in Table 3.

Considering the present results \cite{pdg} of the
experimental searches for the right-handed weak bosons,
we take $\kappa\equiv \Lambda_R/\Lambda_L\geq 10$, so
that we conclude that the allowed regions of $\kappa$,
the intermediate energy scale $\Lambda_S$ and the
unification scale $\Lambda_X\equiv (\Lambda_{XL}
\Lambda_{XL})^{1/2}$ are 
$$
\kappa=1.3\times 10^1 - 1.2\times 10^6 \ ,
$$
$$
\Lambda_S=(6.0\times 10^4 -5.5\times 10^9)\ {\rm GeV}\ ,
\eqno(4.31)
$$
$$
\Lambda_X=(9.8\times 10^{13} -4.9\times 10^{14})\ {\rm GeV}\ ,
$$
corresponding to the values $\alpha_{2R}^{-1}(\Lambda_R)=6-1$.
Behaviors of the gauge coupling constants for a typical case
are illustrated in Fig.~1.


\vspace{.2in}

\section{Conclusions}
\label{sec:5}

\vglue.1in

In conclusion, in order to examine the idea that the extra fermions
$F_R$ and $F_L$ in the universal seesaw mass matrix model, 
together with the conventional three family quarks and leptons 
$f_L$ and $f_R$, are assigned to $(f_L+ F_R^c)\sim (16,1)$ and
$(f_R+ F_L^c)\sim (1,16)$ of ${\rm SO(10)}_L \times {\rm SO(10)}_R$,
we have investigated the evolution of the gauge coupling constants
and intermediate mass scales.
The case A, ${\rm SO(10)}_L \times {\rm SO(10)}_R \rightarrow 
[{\rm SU(5)}\times {\rm U(1)}']_L \times [{\rm SU(5)}\times 
{\rm U(1)}']_R$, is ruled out because the results are inconsistent 
with the observed values of the gauge coupling constants at 
$\mu=m_Z$.
The case B, ${\rm SO(10)}_L \times {\rm SO(10)}_R \rightarrow 
[{\rm SU(2)}\times {\rm SU(2)}'\times {\rm SU(4)}]_L \times 
[{\rm SU(2)}\times {\rm SU(2)}'\times {\rm SU(4)}]_R$, 
is allowed for the intermediate energy scale 
$\Lambda_R\sim (10^{1}-10^{6})$ GeV
if we accept a model with $\Lambda_{XL}\neq \Lambda_{XR}$,
where $\Lambda_{XL}$ and $\Lambda_{XR}$ are the unification
scales of ${\rm SO(10)}_L$ and ${\rm SO(10)}_R$, respectively.
We have obtained the allowed regions $\kappa\simeq 10^1-10^6$,
$\Lambda_S\simeq (6\times 10^4 - 6\times 10^9)$ GeV, and
$\Lambda_X=(\Lambda_{XL}\Lambda_{XR})^{1/2}=(5\times 10^{14}
-10^{14})$ GeV correspondingly to $\alpha_{2R}^{-1}(\Lambda_R)
\simeq 6-1$.

In the case B, since $M_L\sim M_R\sim M_N\sim M_F$ 
($F\neq N$), the case gives effective neutrino mass matrix
$M(\nu_L)\simeq -m_L M_L^{-1} m_L^T$,
so that the conventional neutrino masses $m(\nu_i)$ are of the
order of $m(e_i)/\kappa$.
However, for the condition $\alpha_{2R}^{-1}(\Lambda_R)\geq 1$,
which is a condition that the model is perturbative, the value
of $\kappa$ has been constrained by (4.26), i.e., 
$\kappa\leq 1.20\times 10^6$.
This suggests that $m(\nu_\tau)\sim m(\tau)/\kappa \geq
10^3$ eV.
Such a large value of $m(\nu_\tau)$ is unlikely.
Therefore, the straightforward application of the case B
to the neutrino mass generation scenario is ruled out.

However, the numerical results in Sec.~4 should not 
taken rigidly, because the calculation was done at 
one-loop. 
Moreover, the results are dependent on
the input value $\Lambda_R/\Lambda_S$.
The value $\Lambda_R/\Lambda_S=0.02$ have been quoted from 
Ref.~\cite{KFzp}, where the value was determined from the
observed value of $m_c/m_t$ on the basis of a specific 
model for $m_L$, $m_R$ and $M_F$.
Exactly speaking, the value 0.02 means 
$y_Lv_Ly_Rv_R/y_Sv_S=0.02$, where $y$'s and $v$'s are 
the Yukawa coupling constants and vacuum expectation values,
respectively.
Because of the numerical uncertainty of $y_L$, $y_R$, and
$y_S$, the numerical results may be 
changed by one or two order.
The case B cannot still be ruled out.

In the present paper, the cases for SUSY version of the 
model have not been investigated systematically,
because many versions for the energy scale of the SUSY
partners of the super heavy fermions $F$ can be 
considered.
Nevertheless, the case A can easily be ruled out by 
simple consideration.
On the other hand, for the case B, it is a future task
whether the SUSY version is allowed or not.

When we take the numerical result of the constraint (4.26), 
we can consider a minimum modification of the case B.
In the case B, the Dirac mass matrix $M_D$ is generated
by the Higgs scalar $\Phi_V\sim (1,2,4;1,2,4)$ of
$[{\rm SU(2)}\times {\rm SU(2)}'\times {\rm SU(4)}]_L \times 
[{\rm SU(2)}\times {\rm SU(2)}'\times {\rm SU(4)}]_R$,
while the Majorana mass matrices $M_L$ and $M_R$ are 
generated by the Higgs scalars 
$\Phi_L\sim (1,1,10;1,1,1)$ and $\Phi_R\sim (1,1,1;1,1,10)$,
respectively.
We assume that the symmetries ${\rm SU(4)}_L$ and 
${\rm SU(4)}_R$ are broken into 
$[{\rm SU(3)}\times {\rm U(1)}]_L$ and 
$[{\rm SU(3)}\times {\rm U(1)}]_R$ at 
$\mu=\Lambda_{NL}\equiv O(M_L)$ and 
$\mu=\Lambda_{NR}\equiv O(M_R)$, respectively, and 
the energy scales $\Lambda_{NL}$ and $\Lambda_{NR}$
sufficiently larger than $\Lambda_S\equiv O(M_D)$, 
at which all the fermions $F$ (not $f$) have Dirac masses
$M_F$ and the symmetries ${\rm SU(3)}_L\times {\rm SU(3)}_R$
and ${\rm U(1)}_L\times {\rm U(1)}_R$ are broken into
${\rm SU(3)}_{L+R}\equiv {\rm SU(3)}_c$ and 
${\rm U(1)}_{L+R}\equiv {\rm U(1)}_Y$, respectively.
Then, the neutrino mass generation scenario is changed
from the scenario (b) to the scenario (a).
Although the expression of $M_\nu$ is still given by
$M_\nu\simeq -m_L M_L^{-1}m_L^T$, the suppression 
factor for neutrino masses is changed from $1/\kappa$ 
to $(1/\kappa)(\Lambda_S/\Lambda_{NL})$.
By taking $\Lambda_S/\Lambda_{NL}\sim 10^{-3}$, we can 
obtain reasonable values of the neutrino masses for the
case $\alpha_{2R}^{-1}(\Lambda_R)\simeq 1$.
Of course, in the modified version with 
$\Lambda_{XL}\gg\Lambda_{NL}\gg\Lambda_S$, the 
unification scales of $\Lambda_{XL}$ and 
$\Lambda_{XR}$ are changed by an order of one or two.
However, $\Lambda_R$ and $\Lambda_S$ are insensitive
to the present modification.

In the present paper, we have not discussed the 
evolution of the Yukawa coupling constants.
The phenomenological success in Ref.\cite{KFzp} 
has been obtained by taking $b_e=0$, $b_u=-1/3$
and $b_d=-e^{i\beta_d}$ ($\beta_d=18^\circ$), 
where $M_F=m_0 \lambda_f {\rm diag}(1,1,1+3b_f)$
in the basis on which $M_F$ is diagonal.
The shapes (not the magnitudes) of 
$M_E=m_0 \lambda_e {\rm diag}(1,1,1)$ and
$M_U=m_0 \lambda_u {\rm diag}(1,1,0)$ are
almost invariant under the evolution, while
the shape of $M_D\simeq m_0 \lambda_d {\rm diag}(1,1,-2)$
is not invariant.
The following problems remain as our future tasks:
(i) what value of $b_d$ is favorable at the unification
scale $\mu=\Lambda_X$; 
(ii) whether we can still assert $\lambda_u\simeq \lambda_d$
or not; (iii) whether the mass matrix $m_R$ can still
be approximately diagonal on the basis on which $m_L$ is
diagonal; and so on.
The numerical results in Ref.\cite{KFzp} will be somewhat
cahnged under the present ${\rm SO(10)}_L \times {\rm SO(10)}_R$
model.

In any case, for the universal seesaw mass matrix model 
based on the ${\rm SO(10)}_L \times {\rm SO(10)}_R$ 
unification, if we consider the symmetry breaking 
${\rm SO(10)}_L \times {\rm SO(10)}_R\rightarrow 
[{\rm SU(2)}\times {\rm SU(2)}' \times {\rm SU(4)}]_L
\times 
[{\rm SU(2)}\times {\rm SU(2)}' \times {\rm SU(4)}]_R$
and we accept the case $\Lambda_{XL}\neq \Lambda_R$,
where $\Lambda_{XL}$ and $\Lambda_R$ are the unification
scales of ${\rm SO(10)}_L$ and ${\rm SO(10)}_R$, respectively,
we can find a solution of the intermediate energy scales
$\Lambda_R$ and $\Lambda_S$ for the unified description
of the quark and lepton mass matrices, where only the top
quark mass $m_t$ is given by $m_t\sim \Lambda_L$ in contrast
with $n_q\ll \Lambda_L$ ($q\neq t$) and the neutrino masses
$m(\nu_i)$ are reasonably suppressed compared with the
charged lepton masses $m(e_i)$.
The model is worth while being taken seriously as a 
promising unified model of the quarks and leptons.

\vspace*{.2in}

\centerline{\Large\bf Acknowledgments}

The author would like to thank H.~Fusaoka, T.~Morozumi, 
M.~Tanimoto and T.~Matsuki for  helpful discussions and their 
useful comments.
He also wishes to thank  the Yukawa Institute of Theoretical 
Physics, Kyoto University for its kind hospitality and
offering of the computer facility.
This work was supported by the Grand-in-Aid for Scientific
Research, Ministry of Education, Science and Culture,
Japan (No.~08640386).

\vglue.2in

\newpage

\begin{table}
\caption{ Quantum numbers of the fermions $f$ and $F$
and Higgs scalars $\phi_L$, $\phi_R$ and $\Phi$ for 
${\rm SU(2)}_L\times {\rm SU(2)}_R \times {\rm U(1)}_Y$.
\label{T-qn}}

\begin{tabular}{|c|ccc|c|ccc|} \hline
  & $I_3^L$ & $I_3^R$ & $Y$ &  & $I_3^L$ & $I_3^R$ & $Y$ \\ \hline
$u_L$ & $+\frac{1}{2}$ & 0 & $\frac{1}{3}$ & $u_R$ & 0 & $+\frac{1}{2}$
& $\frac{1}{3}$ \\
$d_L$ & $-\frac{1}{2}$ & 0 & $\frac{1}{3}$ & $d_R$ & 0 & $-\frac{1}{2}$
& $\frac{1}{3}$ \\ \hline
$\nu_L$ & $+\frac{1}{2}$ & 0 & $-1$ & $\nu_R$ & 0 & $+\frac{1}{2}$ & $-1$ \\
$e_L$ & $-\frac{1}{2}$ & 0 & $-1$ & $e_R$ & 0 & $-\frac{1}{2}$ & $-1$ \\ \hline
$U_L$ & 0 & 0 & $\frac{4}{3}$ & $U_R$ & 0 & 0 & $\frac{4}{3}$ \\
$D_L$ & 0 & 0 & $-\frac{2}{3}$ & $D_R$ & 0 & 0 & $-\frac{2}{3}$ \\ \hline
$N_L$ & 0 & 0 & 0 & $N_R$ & 0 & 0 & 0 \\ 
$E_L$ & 0 & 0 & $-2$ & $E_R$ & 0 & 0 & $-2$ \\ \hline
$\phi_L^+$ & $+\frac{1}{2}$ & 0 & 1 & $\phi_R^+$ & 0 & $+\frac{1}{2}$ & 1 \\ 
$\phi_L^0$ & $-\frac{1}{2}$ & 0 & 1 & $\phi_R^0$ & 0 & $-\frac{1}{2}$ & 1 \\ 
\hline
\end{tabular}

\end{table}

\begin{table}
\caption{Coefficients in the evolution equations of 
gauge coupling constants. The cases A and B are 
cases with the symmetry breaking patterns
${\rm SO(10)}\rightarrow{\rm SU(5)}\times {\rm U(1)}$ and
${\rm SO(10)}\rightarrow{\rm SU(2)}\times {\rm SU(2)}
\times{\rm SU(4)} $, which are discussed in Secs.~3
and 4, respectively.\label{T-b}}
\vglue.1in

\begin{tabular}{|c|c|c|c|c|}\hline
      & $\Lambda_L <\mu\leq\Lambda_R$ & 
$\Lambda_R <\mu\leq\Lambda_S$
& \multicolumn{2}{c|}{ $\Lambda_S <\mu\leq\Lambda_X$} 
\\ \cline{4-5}
    &   &   & Case A & Case B \\ \hline
${\rm SU(3)}_c$ & $b^I_3=7$ & $b^{II}_3=19/3$\ (7) & 
$\left\{ \begin{tabular}{c}
$b^{III}_{3L}=6$ \\ 
$b^{III}_{3R}=6$ \\ 
\end{tabular} \right.$
&
$\left\{ \begin{tabular}{c}
$b^{III}_{4L}=7$ \\ 
$b^{III}_{4R}=7$ \\ 
\end{tabular} \right.$ \\ \hline
${\rm SU(2)}_L$ & $b^I_{2L}=19/6$ & $b^{II}_{2L}= 
19/6$\ (19/6) & 
$b^{III}_{2L}=19/6$ & $b^{III}_{2L}=19/6$ \\ \hline
${\rm SU(2)}_R$ &  & $b^{II}_{2R}=19/6$\ (19/6) & 
$b^{III}_{2R}=19/6$ & $b^{III}_{2R}=19/6$  \\ 
\cline{1-1}\cline{3-5}
${\rm U(1)}_Y$ & ${b^I_1}=-41/10$ & $b_1^{II}=-43/6$\ 
($-9/2$) & 
$\left\{ \begin{tabular}{c}
$b^{III}_{1L}=-53/10$ \\ 
$b^{III}_{1R}=-53/10$ \\ 
\end{tabular} \right.$
&
$\left\{ \begin{tabular}{c}
$b^{\prime}_{2L} = -13/6$ \\
$b^{\prime}_{2R} = -13/6$ \\
\end{tabular} \right.$ \\ \hline
\end{tabular}

\end{table}


\begin{table}
\caption{Intermediate mass scales $\Lambda_R$ and $\Lambda_S$ 
versus $\alpha^{-1}_{2R}(\Lambda_R)$ in the case of 
${\rm SO(10)}\rightarrow {\rm SU(2)}\times {\rm SU(2)}
\times {\rm SU(4)}$. Input values 
$\Lambda_R/\Lambda_S=0.02$ and $\Lambda_L=m_Z=91.2$ GeV 
are used.
The upper and lower rows of $\Lambda_{XR}$ and
$\Lambda_{XL}$ correspond to the values for 
$\alpha^{-1}_{4R}(\Lambda_S)=1$ and 
$\alpha^{-1}_{4R}(\Lambda_S)=2$, respectively.
\label{T-a2r}}

\vglue.1in

\begin{tabular}{|c|c|c|c|c|}\hline
$\alpha^{-1}_{2R}(\Lambda_R)$ & 1 & 2 & 4 & 6 \\ \hline
$\Lambda_R/\Lambda_L$ & $1.20 \times 10^6$ &
$1.23\times 10^5$ & $1.27\times 10^3$ & $1.32\times 10^1$ \\ \hline
$\Lambda_R$ [GeV] & $1.10\times 10^8$ & $1.12\times 10^7$ &
$1.16\times 10^5$ & $1.21\times 10^3$ \\ \hline
$\Lambda_S$ [GeV] & $5.48\times 10^{9}$ & $5.59\times 10^8$ &
$5.81\times 10^6$ & $6.04\times 10^4$ \\ \hline
$\Lambda_X$ [GeV] & $4.88\times 10^{14}$ & $3.53\times 10^{13}$ &
$1.86\times 10^{14}$ & $9.75\times 10^{13}$ \\ \hline
$\Lambda_{XR}$ [GeV] &
\begin{tabular}{c}
$1.39 \times 10^{11}$ \\
$2.69 \times 10^{10}$ \\
\end{tabular}
& \begin{tabular}{c}
$7.29 \times 10^{10}$ \\
$1.41 \times 10^{10}$ \\
\end{tabular}
& \begin{tabular}{c}
$2.01 \times 10^{10}$ \\
$3.90 \times 10^{9}$ \\
\end{tabular}
& \begin{tabular}{c}
$5.54 \times 10^{9}$ \\
$1.08 \times 10^9$ \\
\end{tabular} \\ \hline
$\Lambda_{XL}$ [GeV]
& \begin{tabular}{c}
$1.71 \times 10^{18}$ \\
$8.83 \times 10^{18}$ \\
\end{tabular}
& \begin{tabular}{c}
$1.71 \times 10^{18}$ \\
$8.83 \times 10^{18}$ \\
\end{tabular}
& \begin{tabular}{c}
$1.71 \times 10^{18}$ \\
$8.83 \times 10^{18}$ \\
\end{tabular}
& \begin{tabular}{c}
$1.71 \times 10^{18}$ \\
$8.83 \times 10^{18}$ \\
\end{tabular}
 \\ \hline
\end{tabular}

\end{table}


\begin{figure}
\begin{center}
\epsfile{file=so10-g.eps,scale=0.65}
\end{center}
\caption{
Behaviors of $\alpha_1^{\prime -1}(\mu)$ (dotted line) in
$\Lambda_L<\mu\leq \Lambda_R$, $\alpha_1^{-1}(\mu)$
(dotted line) in $\Lambda_R<\mu\leq \Lambda_S$,
$\alpha_{2L}^{-1}(\mu)$ (solid line)
in $\Lambda_L<\mu\leq \Lambda_{XL}$,
$\alpha_{2R}^{-1}(\mu)$ (solid line)
in $\Lambda_R<\mu\leq \Lambda_{XR}$,  
$\alpha_3^{-1}(\mu)$ (dashed line)
in $\Lambda_L<\mu\leq \Lambda_S$,
$\alpha_{2L}^{\prime -1}(\mu)$ (dotted line) 
in $\Lambda_S<\mu\leq \Lambda_{XL}$, and
$\alpha_{2R}^{\prime -1}(\mu)$ (dotted line) 
in $\Lambda_S<\mu\leq \Lambda_{XR}$, 
$\alpha_{4L}^{-1}(\mu)$ (dotted chain line) 
in $\Lambda_S<\mu\leq \Lambda_{XL}$, and
$\alpha_{4R}^{-1}(\mu)$ (dotted chain line) 
in $\Lambda_S<\mu\leq \Lambda_{XR}$, 
where $\Lambda_L=91.2$ GeV, $\Lambda_R=1.10\times 10^{8}$ GeV,
$\Lambda_S=5.48\times 10^{9}$ GeV, 
$\Lambda_{XR}=1.39\times 10^{11}$ GeV and
$\Lambda_{XL}=1.71\times 10^{18}$ GeV.
The values $\alpha_1^{\prime -1}(\Lambda_L)=59.42$,
$\alpha_{2L}^{-1}(\Lambda_L)=29.86$,
$\alpha_3^{-1}(\Lambda_L)=8.410$, 
$\alpha_{2R}^{-1}(\Lambda_R)=1$ and 
$\alpha_{4R}^{-1}(\Lambda_S)=1$ are used as the input values.
}
\end{figure}

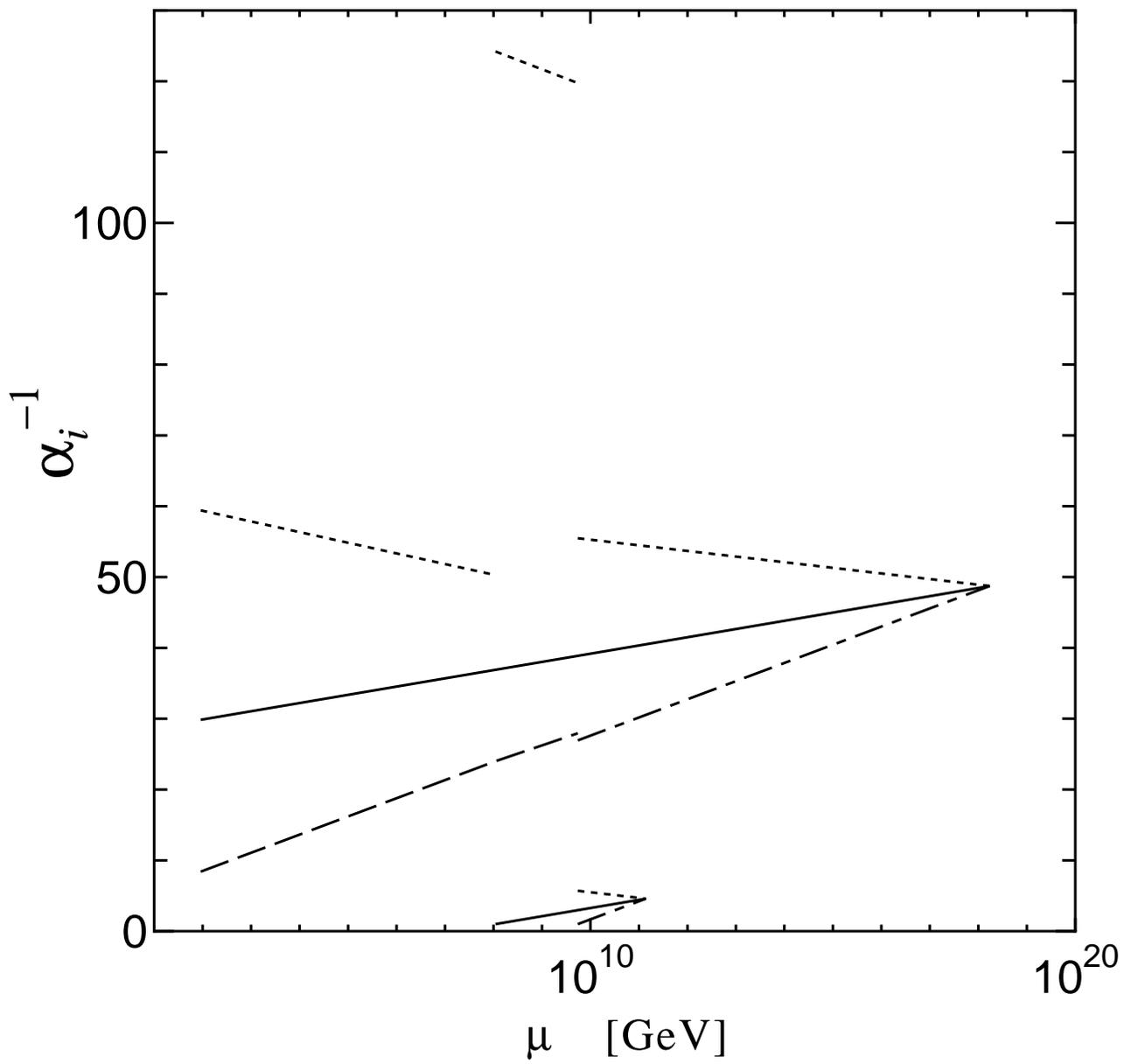
\end{document}